\documentclass[preprint,amssymb,superscriptaddress]{revtex4}
\usepackage{amsmath}
\usepackage{amsthm}
\usepackage{amssymb}
\usepackage{amsfonts}
\usepackage{graphicx}
\usepackage{wasysym}
\usepackage{mathrsfs}
\usepackage{yfonts}
\usepackage{bbold}
\usepackage{verbatim}
\usepackage{subfigure}

\newcommand{\x}{\mathbf{r},t}

\newcommand{\xf}{\mathbf{r},\omega}

\newcommand{\W}{\overleftrightarrow{W}}
\newcommand{\R}{\overleftrightarrow{\mathcal{R}}}
\newcommand{\T}{\overleftrightarrow{\mathcal{T}}}
\renewcommand{\r}{\mathbf{r}}
\newcommand{\E}{\mathbf{E}}

\renewcommand{\k}{\mathbf{k}}
\begin{document}

\title{Effects of Refraction and Reflection on Coherence Properties of Light}

\author{Mayukh Lahiri}
\email{mayukh@pas.rochester.edu} \affiliation{Department of Physics
and Astronomy, University of Rochester, Rochester, NY 14627, U.S.A}

\author{Emil Wolf}
\affiliation{Department of Physics and Astronomy, University of
Rochester, Rochester, NY 14627, U.S.A} \affiliation{Institute of
Optics, University of Rochester, Rochester, NY 14627, U.S.A}


\begin{abstract}
\qquad Partially coherent light beams are encountered both in
classical and in quantum optics. Their coherence properties
generally depend on the correlation properties of their sources. In
this paper, we propose a technique for controlling the coherence
properties of optical beams in laboratory environment. The technique
is based on the fact that coherence properties of partially coherent
electromagnetic beams can be changed on refraction and on
reflection, and that the changes can be controlled by varying the
angle of incidence. \vskip 1cm

\end{abstract}
\maketitle

\section*{Introduction}\label{Sec-int}
Coherence properties of light play important roles in many
experiments both in classical and quantum optics. These properties
are generally determined by the correlation properties of the
source, which generates the light. Change of coherence properties of
light due to propagation (\cite{BW}, Ch. 10) and scattering
(\cite{EW}, Ch. 6) have been studied in great detail. However, such
changes cannot easily be controlled in a laboratory environment. We
propose a method of controlling them in optical experiments. We show
that coherence properties of partially coherent beams can be both
improved and degraded by means of reflection or refraction at a
surface separating two media of different dielectric properties. We
also show that such changes can be controlled by varying the angle
of incidence.
\par
Properties of refracted and reflected light are usually studied by
employing the classic Fresnel formulas. An account of the theory
leading to them can be found, for example, in Ref. \cite{Jackson},
chapter 7. The treatment is based on the assumption that the
incident, the refracted and the transmitted electromagnetic fields
are monochromatic plane waves. The Fresnel formulas, therefore, do
not apply to partially coherent beams, and they cannot provide any
information as to whether coherence properties of light can change
on refraction or on reflection. However, one can use them to
formulate a general theory of refraction and reflection for
partially coherent beams. Application of the generalized formulation
shows that coherence properties of a partially coherent beam change,
in general, on refraction and on reflection.

\section*{Coherence theory of stochastic electromagnetic
beams}\label{sec-coh-th-bas-res} In the optical and in the higher
frequency ranges of the electromagnetic spectrum, the concept of
monochromaticity is an idealization, which is not encountered in
practice. All optical fields exhibit some random fluctuations. If
these fluctuations are assumed to be statistically stationary, the
field can be represented, at each frequency $\omega$, by an ensemble
$\left\{\E(\xf)\right\}$ of monochromatic vector fields (see, for
example, \cite{EW}, Sec. 4.1). When the field is beam-like, one can
neglect the field components along the propagation direction. Hence,
each member of the ensemble of electric field can be represented in
terms of two mutually orthogonal components, each of which is
perpendicular to the direction of propagation. We label them by the
symbols v and h. Each member of the ensemble of the electric field
vectors can be represented as a column matrix, i.e., in the form
\begin{equation}\label{E-pl-wv-mat}
\E(\xf)=
\begin{pmatrix} E_{\text{v}}(\xf)
\\
E_{\text{h}}(\xf)
\end{pmatrix}=
\begin{pmatrix} E_{\text{v}}(\xf)
& \quad E_{\text{h}}(\xf)
\end{pmatrix}^T,
\end{equation}
where the superscript $T$ denotes the transpose of the matrix. The
second-order correlation properties of such a field
\cite{Note-quant-coh-order} are characterized by a $2\times 2$
correlation matrix $\W(\r_1,\r_2;\omega)$, the so-called
cross-spectral density matrix (CSDM), which is defined, at a pair of
points specified by position vectors $\mathbf{r}_1$ and
$\mathbf{r}_2$, by the formula (\cite{EW}, Ch. 9)
\begin{equation}\label{CSDM-re-wr}
\W(\r_1,\r_2;\omega)= \left \langle \E^{\ast}(\r_1;\omega) \cdot
\E^T(\r_2;\omega) \right \rangle.
\end{equation}
Here the asterisk denotes the complex conjugate, the dot denotes
matrix multiplication, and the angular brackets denote the ensemble
average. Clearly a typical element of the CSDM is
$W_{lm}(\r_1,\r_2;\omega)=\langle
E_{\text{l}}^{\ast}(\mathbf{r}_1;\omega)E_{\text{m}}(\mathbf{r}_2;\omega)\rangle$;
$l$=h,v; $m$=h,v.
\par
Coherence properties of an optical beam characterize its ability to
interfere. The simplest types of coherence properties are
characterized by the visibility of fringes produced in an Young's
interference experiment. Suppose that a light beam is incident on an
opaque screen containing two pinholes located at points $\r_1$ and
$\r_2$. The visibility of the interference fringes produced by the
frequency component $\omega$ at another screen placed sufficiently
far behind the pinholes, is given by the modulus
$|\eta(\r_1,\r_2;\omega)|$ of the spectral degree of coherence,
i.e., of the spatial degree of coherence at frequency $\omega$,
defined by the formula (\cite{EW}, Ch. 9)
\begin{equation}\label{doc-em-def}
\eta(\r_1,\r_2;\omega) \equiv \frac{\text{Tr}~\W(\r_1,\r_2;\omega)}
{\sqrt{\text{Tr}~\W(\r_1,\r_1;\omega)}\sqrt{\text{Tr}~\W(\r_2,\r_2;\omega)}},
\end{equation}
where Tr $\W$ denotes the trace of the matrix $\W$. It can readily
be shown that $0\leq |\eta(\r_1,\r_2;\omega)|\leq 1$. When
$|\eta(\r_1,\r_2;\omega)|=1$, i.e., when the fringe-visibility is
maximum, the beam is said to be spatially completely coherent at the
pair of points $(\r_1,\r_2)$. In the other extreme case when
$\eta(\r_1,\r_2;\omega)=0$, the beam is said to be spatially
incoherent at the two points. In any intermediate case ($0 <
|\eta(\r_1,\r_2;\omega)| < 1$), the beam is said to be partially
coherent at the two points, at frequency $\omega$.
\par
Another definition of the degree of coherence of electromagnetic
beams have been proposed (for some discussions relating to this
topic see \cite{STF1,W-comment,STF1-re-comment}). For our purpose,
it is immaterial which of the two definitions is used. In this
paper, we use the definition in terms of fringe visibility, because
it is often employed in the analysis of experimental results (see,
for example, \cite{Laura-coh-synth}).

\section*{Fresnel formulas for reflection and refraction of
monochromatic plane waves} \label{subsec-fres-formulas} Let us first
consider refraction and reflection of a monochromatic plane wave at
a planar interface that separates two homogeneous media. Suppose
that dielectric properties of the two media are characterized by
permittivities and permeabilities $\epsilon$, $\mu$, and
$\epsilon'$, $\mu'$. Their refractive indices are given by
$n=\sqrt{\epsilon \mu / \epsilon_0 \mu_0}$, and $n'=\sqrt{\epsilon'
\mu' / \epsilon_0 \mu_0}$, respectively (see, for example,
\cite{Jackson}, p. 303). Here $\epsilon_0\approx 8.854\times
10^{-12}$ F$/$m is the vacuum permittivity, and $\mu_0\approx
1.257\times 10^{-6}$ H$/$m is the vacuum permeability.
\par
Suppose that a monochromatic plane wave is incident on the interface
at an angle of incidence $\theta_{\mathbb{i}}$ (see Fig.
\ref{fig:pl-wv-in-tr-ref-illus}). The electric field vector of the
incident wave can be expressed in the form
$\E^{(\mathbb{i})}(\x)=\E_0^{(\mathbb{i})}\exp\left[i(\k^{(\mathbb{i})}\cdot
\r-\omega t )\right],$ where $\k^{(\mathbb{i})}$ is the wave vector.
Similarly, the transmitted and the reflected electric field vectors
may be represented by the expressions
$\E^{(\mathbb{t})}(\x)=\E_0^{(\mathbb{t})}\exp\left[i(\k^{(\mathbb{t})}\cdot
\r-\omega t )\right],$ and
$\E^{(\mathbb{r})}(\x)=\E_0^{(\mathbb{r})}\exp\left[i(\k^{(\mathbb{r})}\cdot
\r-\omega t )\right],$ respectively. The moduli of the wave vectors
are given by the formulas $\left| \k^{(\mathbb{t})}
\right|=\omega\sqrt{\epsilon'\mu'}$, and $\left| \k^{(\mathbb{i})}
\right|=\left| \k^{(\mathbb{r})} \right| =\omega\sqrt{\epsilon\mu}$
[see, for example, \cite{Jackson}, Eq. (7.33)]. The plane formed by
the wave vector $\k^{(\mathbb{i})}$ and the normal $\mathbf{n}$ to
the interface defines the plane of incidence. The refracted and the
reflected wave vectors also lie in this plane.
\par
Because electromagnetic waves are transverse, there is no component
of the electric field vector in the direction of propagation of the
incident plane wave. Hence,  $\E^{(\mathbb{i})}$ can be expressed in
terms of two mutually orthogonal components
$\E_{\text{v}}^{(\mathbb{i})}$ and $\E_{\text{h}}^{(\mathbb{i})}$,
i.e., $\E^{(\mathbb{i})}(\x)=\E_{\text{v}}^{(\mathbb{i})}(\x)
+\E_{\text{h}}^{(\mathbb{i})}(\x)=\left(\E_{0\text{v}}^{(\mathbb{i})}
+\E_{0\text{h}}^{(\mathbb{i})}\right)\exp\left[i(\k^{(\mathbb{i})}
\cdot \r-\omega t )\right]$. We choose
$\E_{\text{v}}^{(\mathbb{i})}$ and $\E_{\text{h}}^{(\mathbb{i})}$ to
be perpendicular and parallel, respectively, to the plane of
incidence (see Fig. \ref{fig:pl-wv-in-tr-ref-illus}). Similarly the
transmitted and the reflected electric fields can be uniquely
decomposed in the v and the h directions, i.e., one has
$\E^{(\mathbb{t})}(\x)=\left(\E_{0\text{v}}^{(\mathbb{t})}
+\E_{0\text{h}}^{(\mathbb{t})}\right)\exp\left[i(\k^{(\mathbb{t})}\cdot
\r-\omega t )\right]$, and
$\E^{(\mathbb{r})}(\x)=\left(\E_{0\text{v}}^{(\mathbb{r})}
+\E_{0\text{h}}^{(\mathbb{r})}\right)\exp\left[i(\k^{(\mathbb{r})}
\cdot \r-\omega t )\right],$ respectively.
\par
At the interface, the components of the transmitted and of the
reflected fields are related to the components of the incident field
by the well known Fresnel formulas, which can be expressed in the
matrix form as
\begin{equation}\label{Fresnel-rels-mat}
\E_0^{(\mathbb{t})}=\overleftrightarrow{\text{T}} \cdot
\E_0^{(\mathbb{i})}, \qquad
\E_0^{(\mathbb{r})}=\overleftrightarrow{\text{R}} \cdot
\E_0^{(\mathbb{i})}.
\end{equation}
Here $\overleftrightarrow{\text{T}}=\bigl(
\begin{smallmatrix} T_{\text{v}} & 0
\\
0 & T_{\text{h}}
\end{smallmatrix}\bigr)$ and
$\overleftrightarrow{\text{R}}=\bigl(
\begin{smallmatrix} R_{\text{v}} & 0
\\
0 & R_{\text{h}} \end{smallmatrix}\bigr)$ are two diagonal matrices,
whose elements are given by [see, for example, \cite{Jackson}, Eqs.
(7.39), and (7.41)]
\begin{subequations}\label{Fresnel-rels-pl-wv}
\begin{align}
T_{\text{v}}&=\frac{2n\cos\theta_{\mathbb{i}}}{n\cos\theta_{\mathbb{i}}
+\frac{\mu}{\mu'}\sqrt{n'^2-n^2\sin^2\theta_{\mathbb{i}}}}, \qquad
T_{\text{h}}=\frac{2n n'
\cos\theta_{\mathbb{i}}}{\frac{\mu}{\mu'}n'^2\cos\theta_{\mathbb{i}}
+n\sqrt{n'^2-n^2\sin^2\theta_{\mathbb{i}}}}, \label{Fresnel-rels-pl-wv:a} \\
R_{\text{v}}&=\frac{n\cos\theta_{\mathbb{i}}-\frac{\mu}{\mu'}
\sqrt{n'^2-n^2\sin^2\theta_{\mathbb{i}}}}{n\cos\theta_{\mathbb{i}}
+\frac{\mu}{\mu'}\sqrt{n'^2-n^2\sin^2\theta_{\mathbb{i}}}}, \qquad
R_{\text{h}}=\frac{\frac{\mu}{\mu'}n'^2\cos\theta_{\mathbb{i}}-n\sqrt{n'^2-n^2\sin^2\theta_{\mathbb{i}}}}
{\frac{\mu}{\mu'}n'^2\cos\theta_{\mathbb{i}}+n\sqrt{n'^2-n^2\sin^2\theta_{\mathbb{i}}}},
\label{Fresnel-rels-pl-wv:b}
\end{align}
\end{subequations}
$\theta_{\mathbb{i}}$ being the angle of incidence.

\section*{Theory of refraction and reflection with partially coherent beams}\label{sec-par-coh-refl-refr}
While dealing with partially coherent beams, one does not have the
simplicity associated with monochromatic plane waves. One must then
consider the effects of random fluctuations that are present in the
electric field. Moreover, since partially coherent beams are not
plane waves, one needs to employ the angular spectrum representation
to study their properties (for a basic description of angular
spectrum representation see, for example, \cite{MW}, Sec. 3.2).
According to the theory relating to such a representation, a
partially coherent beam may be represented as superposition of
ensembles of homogeneous and of evanescent plane waves. However, if
one does not consider total internal reflection, the contribution
from evanescent waves can usually be neglected
\cite{Note-tot-int-refl}. On refraction (reflection) of a partially
coherent incident beam, the plane wave components present in it are
individually refracted (reflected). Each of the plane wave
components has a different direction of propagation, and, therefore,
their angles of incidence are, in general, different from each
other. Hence, the Fresnel coefficients have different values for
each plane wave component. After refraction and reflection the plane
waves will recombine to generate the transmitted and the reflected
partially coherent beams. Because each plane wave undergo
transformation characterized by different Fresnel coefficients, the
beams generated by their recombination after refraction and
reflection have coherence properties which are, in general,
different from those of the incident beam.
\par
The phenomenon can readily be described mathematically. It is useful
to introduce separate coordinate systems for the incident, for the
transmitted, and for the reflected beams. Let us denote them by
$(x_{\text{v}}^{(\mathbb{i})},x_{\text{h}}^{(\mathbb{i})},x_{\text{p}}^{(\mathbb{i})})$,
$(x_{\text{v}}^{(\mathbb{t})},x_{\text{h}}^{(\mathbb{t})},x_{\text{p}}^{(\mathbb{t})})$,
and
$(x_{\text{v}}^{(\mathbb{r})},x_{\text{h}}^{(\mathbb{r})},x_{\text{p}}^{(\mathbb{r})})$
respectively. We define each of them as follows: We choose the
positive direction of the $x_{\text{p}}^{(l)}$ \textendash axis
($l=\mathbb{i},\mathbb{t},\mathbb{r}$) along the axis of the beam in
the direction of propagation, the positive $x_{\text{v}}^{(l)}$
\textendash axis to point at right angle into the plane defined by
$\mathbf{n}$ and the axis of the beam; and the $x_{\text{h}}^{(l)}$
\textendash axis is chosen following the right-hand rule. Figure
\ref{fig:in-beam-coord-sys} illustrates this for the incident beam.
We note that in all the three coordinate systems the v \textendash
directions are the same, i.e., that
$x_{\text{v}}^{(\mathbb{i})}\equiv x_{\text{v}}^{(\mathbb{t})}\equiv
x_{\text{v}}^{(\mathbb{r})}$. It is evident that the coordinate
systems for the three beams are related to each other by
two-dimensional rotations around the v direction. The angles of
incidence ($\theta_{\mathbb{i}}$), of refraction
($\theta_{\mathbb{t}}$), and of reflection ($\theta_{\mathbb{r}}$)
are the angles between the normal $\mathbf{n}$ and the respective
beam axes.
\par
In the angular spectrum representation, the CSDM of a partially
coherent beam is expressed in the form (\cite{MW}, Sec. 5.6.3;
\cite{Note-calc-illus})
\begin{equation}\label{CSDM-in-ang-spec}
\W^{(l)}(\r,\r';\omega)=\iint
\W_A^{(l)}(\mathbf{k}^{(l)}_{\perp},\mathbf{k}^{(l)'}_{\perp};\omega)\exp\left[
i(\mathbf{k}^{(l)'}\cdot \r'-\mathbf{k}^{(l)}\cdot \r)
\right]~d^2k^{(l)}_{\perp}d^2k^{(l)'}_{\perp},
\end{equation}
where the superscript $l$ may represent an incident ($\mathbb{i}$),
a refracted ($\mathbb{t}$), or a reflected ($\mathbb{r}$) beam. The
angular correlation matrix
$\W_A^{(l)}(\mathbf{k}^{(l)}_{\perp},\mathbf{k}^{(l)'}_{\perp};\omega)$
is the cross-correlation matrix formed by the space-independent
parts of field components of two plane waves with wave vectors
$\mathbf{k}^{(l)}$ and $\mathbf{k}^{(l)'}$; and
$\mathbf{k}^{(l)}_{\perp}$ and $\mathbf{k}^{(l)'}_{\perp}$ are
two-dimensional vectors representing the transverse components of
$\mathbf{k}^{(l)}$ and $\mathbf{k}^{(l)'}$ respectively. The
integrations extend over the domains
$|\mathbf{k}^{(l)}_{\perp}|^2,|\mathbf{k}^{(l)'}_{\perp}|^2 \ll
|\k^{(l)}|^2$. In the coordinate system
$(x_{\text{v}}^{(l)},x_{\text{h}}^{(l)},x_{\text{p}}^{(l)})$, the
vectors $\mathbf{k}^{(l)}$ and  $\mathbf{k}^{(l)}_{\perp}$ are
represented by $\mathbf{k}^{(l)}\equiv
(k_{\text{v}}^{(l)},k_{\text{h}}^{(l)},k_{\text{p}}^{(l)})$ and
$\mathbf{k}^{(l)}_{\perp}\equiv
(k_{\text{v}}^{(l)},k_{\text{h}}^{(l)})$. Assuming that the plane
wave component characterized by the wave vector
$\mathbf{k}^{(\mathbb{i})}$ makes an angle of incidence
$\tilde{\theta_{\mathbb{i}}}$, one can show that
$\cos\tilde{\theta_{\mathbb{i}}} \approx
\cos\theta_{\mathbb{i}}+(k_{\text{h}}^{(\mathbb{i})}/|\mathbf{k}^{(\mathbb{i})}|)
\sin\theta_{\mathbb{i}}$, and $\sin\tilde{\theta_{\mathbb{i}}}
\approx
\sin\theta_{\mathbb{i}}-(k_{\text{h}}^{(\mathbb{i})}/|\mathbf{k}^{(\mathbb{i})}|)
\cos\theta_{\mathbb{i}}$. When this plane wave component is
refracted, the angle of refraction $\tilde{\theta_{\mathbb{t}}}$ is
obtained by applying Snell's law
$\sin\tilde{\theta_{\mathbb{t}}}/\sin\tilde{\theta_{\mathbb{i}}}=n/n'
=\sqrt{\mu\epsilon/\mu'\epsilon'}$; and when the plane wave is
reflected, its angle of reflection
$\tilde{\theta_{\mathbb{r}}}=\tilde{\theta_{\mathbb{i}}}$. The plane
of incidence of this wave component is the plane formed by the
vector $\mathbf{k}^{(\mathbb{i})}$ and the normal $\mathbf{n}$ to
the interface. This plane is different from the plane defined by the
axis of the incident beam and the normal $\mathbf{n}$. The former
(A, say) can be obtained by a rotation of the latter (B, say)
through an angle
$\alpha=\tan^{-1}[-k_{\text{v}}^{(\mathbb{i})}/(k_{\text{h}}^{(\mathbb{i})}
\cos\theta_{\mathbb{i}}
-k_{\text{p}}^{(\mathbb{i})}\sin\theta_{\mathbb{i}})]$ around the
normal $\mathbf{n}$ (see Fig. \ref{fig:xyz-z'y'z'-trans}). For each
of the plane wave components, one can now define a coordinate system
for which the v, the h and the p directions are defined in a similar
way to that illustrated in Fig. \ref{fig:pl-wv-in-tr-ref-illus}.
These directions are different for each plane wave component and we
denote these coordinate systems by
$(x_{\text{v}'}^{(l)},x_{\text{h}'}^{(l)},x_{\text{p}'}^{(l)})$,
$l=\mathbb{i},\mathbb{t},\mathbb{r}$.
\par
It is convenient to evaluate the integral in Eq.
(\ref{CSDM-in-ang-spec}) for the incident, for the refracted and for
the reflected beams in their respective coordinate systems. The
matrices
$\W_A^{(\mathbb{t})}(\mathbf{k}^{(\mathbb{t})}_{\perp},\mathbf{k}^{(\mathbb{t})'}_{\perp};\omega)$
and
$\W_A^{(\mathbb{r})}(\mathbf{k}^{(\mathbb{r})}_{\perp},\mathbf{k}^{(\mathbb{r})'}_{\perp};\omega)$
are related to the matrix
$\W_A^{(\mathbb{i})}(\mathbf{k}^{(\mathbb{i})}_{\perp},\mathbf{k}^{(\mathbb{i})'}_{\perp};\omega)$
by the formulas
\begin{equation} \label{WA-trns-refl}
\W_A^{(\mathbb{t})}=\overleftrightarrow{\mathscr{U}}_{\mathcal{T}}^{\ast}
\cdot \W_A^{(\mathbb{i})} \cdot
\overleftrightarrow{\mathscr{U}}_{\mathcal{T}}^{T}, \quad \text{and}
\quad
\W_A^{(\mathbb{r})}=\overleftrightarrow{\mathscr{U}}_{\mathcal{R}}^{\ast}
\cdot \W_A^{(\mathbb{i})} \cdot
\overleftrightarrow{\mathscr{U}}_{\mathcal{R}}^{T}.
\end{equation}
Here the matrices $\overleftrightarrow{\mathscr{U}}_{\mathcal{T}}$
and $\overleftrightarrow{\mathscr{U}}_{\mathcal{R}}$ are different
for each plane wave component present in the angular spectrum of the
beam and, therefore, cannot be treated as constant factors while
performing the integration in Eq. (\ref{CSDM-in-ang-spec}). They can
be represented in the following product forms:
$\overleftrightarrow{\mathscr{U}}_{\mathcal{T}}
=\left\{\overleftrightarrow{\mathscr{U}}^{(\mathbb{t})}
\right\}^{\dag} \cdot \T \cdot
\overleftrightarrow{\mathscr{U}}^{(\mathbb{i})}$, and
$\overleftrightarrow{\mathscr{U}}_{\mathcal{R}}
=\left\{\overleftrightarrow{\mathscr{U}}^{(\mathbb{r})}
\right\}^{\dag} \cdot \R \cdot
\overleftrightarrow{\mathscr{U}}^{(\mathbb{i})}$. The matrices $\T
\equiv \bigl(
\begin{smallmatrix} \mathcal{T}_{\text{v}'} & 0
\\
0 & \mathcal{T}_{\text{h}'}
\end{smallmatrix}\bigr)$ and $\R\equiv \bigl(
\begin{smallmatrix} \mathcal{R}_{\text{v}'} & 0
\\
0 & \mathcal{R}_{\text{h}'}
\end{smallmatrix}\bigr)$ are similar to the Fresnel transformation
matrices for refraction $\overleftrightarrow{\text{T}}$ and
reflection $\overleftrightarrow{\text{R}}$ respectively, defined at
the end of the previous section; however, their elements
$\mathcal{T}_{\text{v}'}$, $\mathcal{T}_{\text{h}'}$,
$\mathcal{R}_{\text{v}'}$ and $\mathcal{R}_{\text{h}'}$ are now
given by expressions that are obtained by replacing
$\theta_{\mathbb{i}}$ by $\tilde{\theta_{\mathbb{i}}}$ in
expressions (\ref{Fresnel-rels-pl-wv}) of $T_{\text{v}}$,
$T_{\text{h}}$, $R_{\text{v}}$ and $R_{\text{h}}$, respectively.
This is so because the Fresnel formulas apply separately for each of
the plane wave components, which has a unique plane of incidence
(A), defined by $\mathbf{k}^{(\mathbb{i})}$ and $\mathbf{n}$. The
matrices $\overleftrightarrow{\mathscr{U}}^{(l)}$,
$(l=\mathbb{i},\mathbb{t},\mathbb{r})$, define the relations between
the v and the h components of the respective fields with their
$\text{v}'$ and $\text{h}'$ components (the p and the $\text{p}'$
components may be neglected for beam-like fields). Their explicit
forms are given by $\overleftrightarrow{\mathscr{U}}^{(\mathbb{i})}
=\left\{\overleftrightarrow{\mathscr{U}}^{(\mathbb{r})}\right\}^{\dag}
=\cos\theta_{\mathbb{i}}
\bigl(\begin{smallmatrix} \cos\alpha /\cos\theta_{\mathbb{i}} & \quad \sin\alpha \\
-\sin\alpha & \quad \cos\alpha \cos\tilde{\theta_{\mathbb{i}}}+
\sin\tilde{\theta_{\mathbb{i}}} \tan\theta_{\mathbb{i}}
\end{smallmatrix}\bigr)$ and
$\overleftrightarrow{\mathscr{U}}^{(\mathbb{t})}=\cos\theta_{\mathbb{t}}
\bigl(\begin{smallmatrix} \cos\alpha /\cos\theta_{\mathbb{t}} & \quad \sin\alpha \\
-\sin\alpha & \quad \cos\alpha \cos\tilde{\theta_{\mathbb{t}}}+
\sin\tilde{\theta_{\mathbb{t}}} \tan\theta_{\mathbb{t}}
\end{smallmatrix}\bigr)$. The dependence on $\alpha$, $\tilde{\theta_{\mathbb{i}}}$
and $\tilde{\theta_{\mathbb{t}}}$ clearly shows that the matrices
$\T$, $\R$, $\overleftrightarrow{\mathscr{U}}^{(\mathbb{t})}$ and
$\overleftrightarrow{\mathscr{U}}^{(\mathbb{r})}$ are, in general,
different for each plane wave component.

\section*{Controlling coherence properties of a beam}
We will now show that if a light beam generated by a partially
coherent source is refracted and reflected, its coherence properties
can change appreciably, and that the change depends on the angle of
incidence. For this purpose, we consider a partially coherent light
source with known correlation properties, i.e., with known CSDM. It
follows from Eq. (\ref{CSDM-in-ang-spec}) that for the incident
beam, the angular correlation matrix
$\W_A^{(\mathbb{i})}(\mathbf{k}^{(\mathbb{i})}_{\perp},\mathbf{k}^{(\mathbb{i})'}_{\perp};\omega)$
is the Fourier transform of the CSDM at the source plane (see
\cite{MW}, Sec. 5.6.3; \cite{Note-calc-illus}). Following the
procedure discussed in the previous section, one can now determine
the CSDMs, and hence the coherence properties of the refracted and
of the reflected beams.
\par
Let us consider a light beam generated by a Gaussian Schell-model
source (see, for example, \cite{GSPBMS,SKW}; see also \cite{EW},
Sec. 9.4.2). Elements of the CSDM of a such light beam, at a pair of
points $(\pmb{\rho}_0,\pmb{\rho}_0')$ in the source plane, can be
expressed in the form
$W_{\beta\gamma}(\pmb{\rho}_0,\pmb{\rho}_0';\omega)=
A_{\beta}A_{\gamma}B_{\beta\gamma}\exp\left[-(\rho^2_0+\rho_0'^2)/(4\sigma^2)
\right]\exp\left[ -(\pmb{\rho}_0'-\pmb{\rho}_0)^2/(2\delta^2)
\right]$, where $\beta=$h,v, and $\gamma=$h,v. The parameters
$A_{\beta}$, $B_{\beta\gamma}$, $\sigma$, and $\delta$ are position
independent. The parameters $B_{\beta\gamma}$, $\sigma$ and $\delta$
cannot be chosen arbitrarily and, in our case, the following
relations must hold (see, for example, \cite{KSW,RK,GSBR}):
$B_{\beta\gamma}=B^{\ast}_{\beta\gamma}$; $B_{\beta\gamma}=1$, when
$\beta=\gamma$; $|B_{\beta\gamma}|\leq 1$, when $\beta \neq \gamma$;
and $1/4\sigma^2 +1/\delta^2 \ll 2\pi^2/\lambda^2$. We take $\omega
\approx 3.2\times 10^{15}$ sec$^{-1}$, and choose the other
parameters as follows: $\delta=0.001$m, $\sigma=0.01$m,
$B_{\text{hv}}=9/16$, and $A_{\text{h}}/A_{\text{v}}=1$ (individual
values of $A_{\text{h}}$ and $A_{\text{v}}$ are not required in the
calculation of the degree of coherence). Suppose now that the beam
generated by this source propagates a distance of 1 meter through
air (refractive index $n\approx 1$), and is then incident on a
planar surface of another medium of refractive index $n'$. Coherence
properties of the reflected and of the transmitted beams are
determined by using Eqs. (\ref{doc-em-def}) and
(\ref{CSDM-in-ang-spec}) for three different media: ethanol
($n'\approx 1.36$), flint glass ($n'\approx 1.62$), and diamond
($n'\approx 2.42$).
\par
We first calculate the moduli of the degrees of coherence of the
transmitted and the reflected beams, at a fixed pair of points on
the interface, for different values of the angle of incidence. One
of the points is chosen as the point of intersection of the incident
beam axis with the interface; and the other point is taken to be
located on the $y$ axis (see Fig. \ref{fig:in-beam-coord-sys}), at a
distance 0.001m away from the first point. Figure
\ref{fig:trans-doc-th-dep} shows the dependence of the modulus of
the degree of coherence of the transmitted beam on the angle of
incidence (plotted up to $89^{\circ}$) for the three media
considered. Starting from a value of $|\eta| \approx 0.61$ at
$\theta_{\mathbb{i}}=0^{\circ}$, the modulus of the degree of
coherence attains its maximum value $|\eta|\approx 1$, at
$\theta_{\mathbb{i}}\approx 56.17^{\circ}$ for ethanol, at
$\theta_{\mathbb{i}}\approx 50.77^{\circ}$ for flint glass, and at
$\theta_{\mathbb{i}}\approx 44.04^{\circ}$ for diamond. Then its
value gradually decreases with increasing angle of incidence and,
eventually, at $\theta_{\mathbb{i}}=89^{\circ}$ it attains values
$|\eta| \approx 0.42$ for ethanol, $|\eta| \approx 0.31$ for flint
glass, and $|\eta| \approx 0.20$ for diamond. We see from this
figure that the coherence properties of the transmitted beam can be
both improved and degraded by varying the angle of incidence.
\par
Figure \ref{fig:refl-doc-th-dep} shows the dependence of the modulus
of the degree of coherence of the reflected beam on the angle of
incidence. It is evident that in this case also the coherence
properties can be both improved and degraded by changing the angle
of incidence. However, in the case of reflection, the dependence is
found to be the same for all the three media, which was not the case
for the transmitted beam. Starting from a value of $|\eta| \approx
0.61$ at $\theta_{\mathbb{i}}=0^{\circ}$, the modulus of the degree
of coherence of the reflected beam first increases gradually to
attain a maximum value $|\eta|\approx 1$ at
$\theta_{\mathbb{i}}\approx 24.09^{\circ}$; then it gradually
decreases to a minimum value $|\eta|\approx 0.21$ at
$\theta_{\mathbb{i}}\approx 58.19^{\circ}$, and finally again
increases with the angle of incidence.
\par
We will now examine the changes in the coherence properties of the
transmitted and the reflected beams at the interface, by varying the
pair of points $(\r,\r')$, for which the degree of coherence would
be determined. We choose the point $\r$ as the point of intersection
of the incident beam axis with the interface, and take the point
$\r'$ as a variable point along the $y$ axis ( $|\r'-\r|=\rho$,
say). In Figs. \ref{subfig:doc-trans-th-56-89-ethanol},
\ref{subfig:doc-trans-th-50-89-flint-glass} and
\ref{subfig:doc-trans-th-44-89-diamond}, the moduli of the degree of
coherence of the transmitted beams are plotted as functions of
$\rho$, for ethanol ($n'\approx 1.36$), for flint glass ($n'\approx
1.62$) and for diamond ($n'\approx 2.42$). In each case, they are
plotted for two different values of the angle of incidence, at which
their minimum and their maximum values were obtained from previous
calculations (cf. Fig. \ref{fig:trans-doc-th-dep}). In each of these
figures, the modulus of the degree of coherence at the source plane
is also plotted as a reference line to display the amount of
controllable change in degree of coherence that can be achieved in
this process.
\par
In Fig. \ref{fig:doc-refl-th-25-50-flint-glass}, the modulus of the
degree of coherence of the reflected beam is plotted as a function
of $\rho$, for two different values of the angle of incidence, at
which its minimum and maximum values were obtained (cf. Fig.
\ref{fig:refl-doc-th-dep}). The modulus of the degree of coherence
at the source plane is also plotted to give an indication of the
amount of controllable change in the degree of coherence that can be
achieved by reflection. It is to be noted that, as regards to its
coherence properties, the reflected light behaves in the same way
for all three types of interfaces.

\section*{Summary}
The fact that coherence properties of light beams can be controlled
by reflecting and refracting them at suitable angles, does not
appear to have been previously noted. This is because the laws of
refraction and reflection for partially coherent light have not been
previously studied. Our results show that it is possible to improve
and to degrade coherence properties of a light beam by refraction or
reflection, and that the change can be controlled by varying the
angle of incidence.

\section*{Acknowledgements}
The research was supported by the US Air Force Office of Scientific
Research under grant No. FA9550-08-1-0417.

\newpage
\begin{center}
\section*{Figure Legends}
\end{center}
\qquad

\begin{figure}[htbp]  \centering
  \includegraphics[scale=.8]{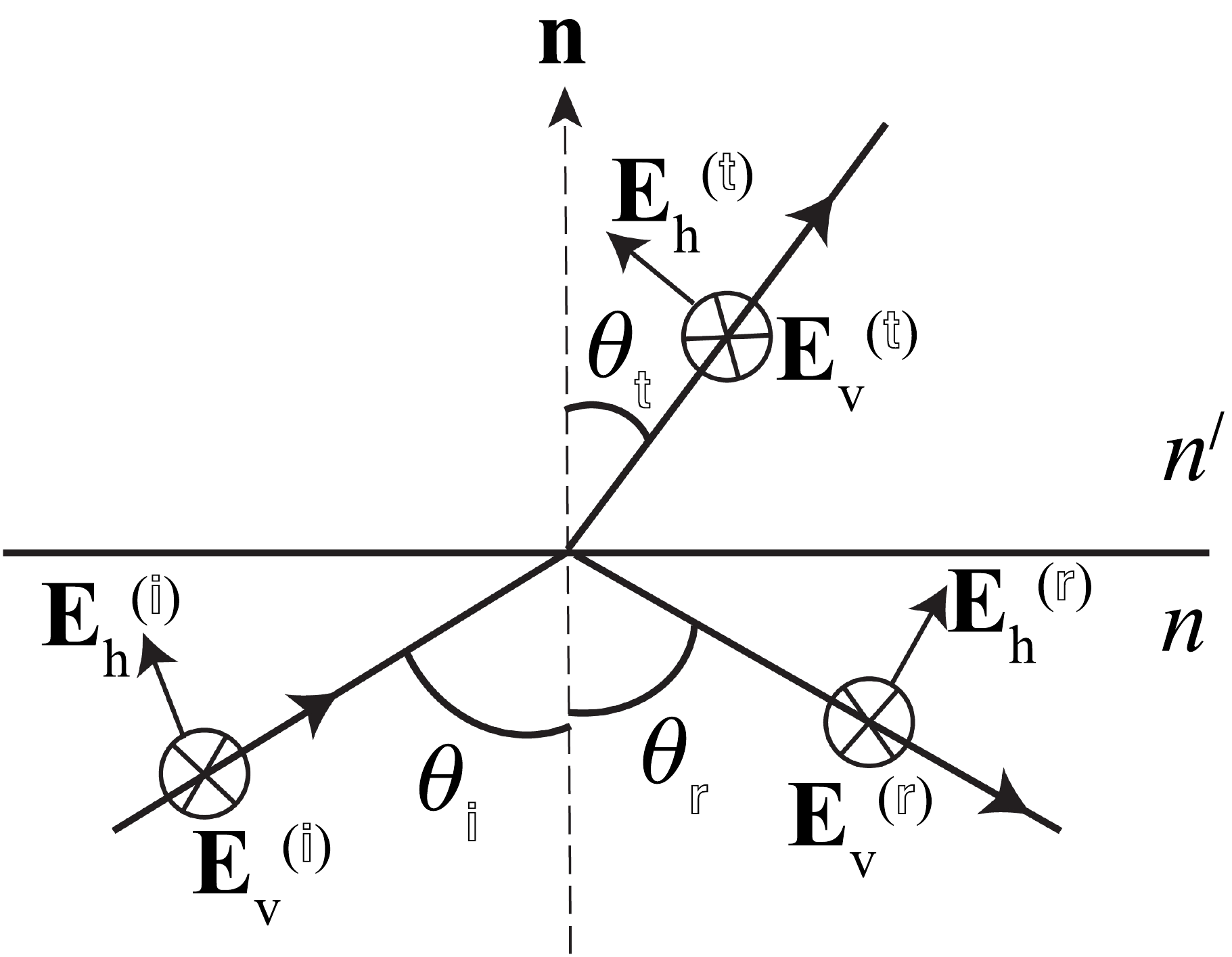}
  \qquad
  \caption{} \label{fig:pl-wv-in-tr-ref-illus}
    \end{figure}

\qquad Fig. \ref{fig:pl-wv-in-tr-ref-illus}. Illustrating the
geometry relating to refraction and reflection of a monochromatic
plane wave at an interface; h and v directions are chosen to be
parallel and perpendicular to the plane of incidence.

\begin{figure}[htbp]  \centering
  \includegraphics[scale=.8]{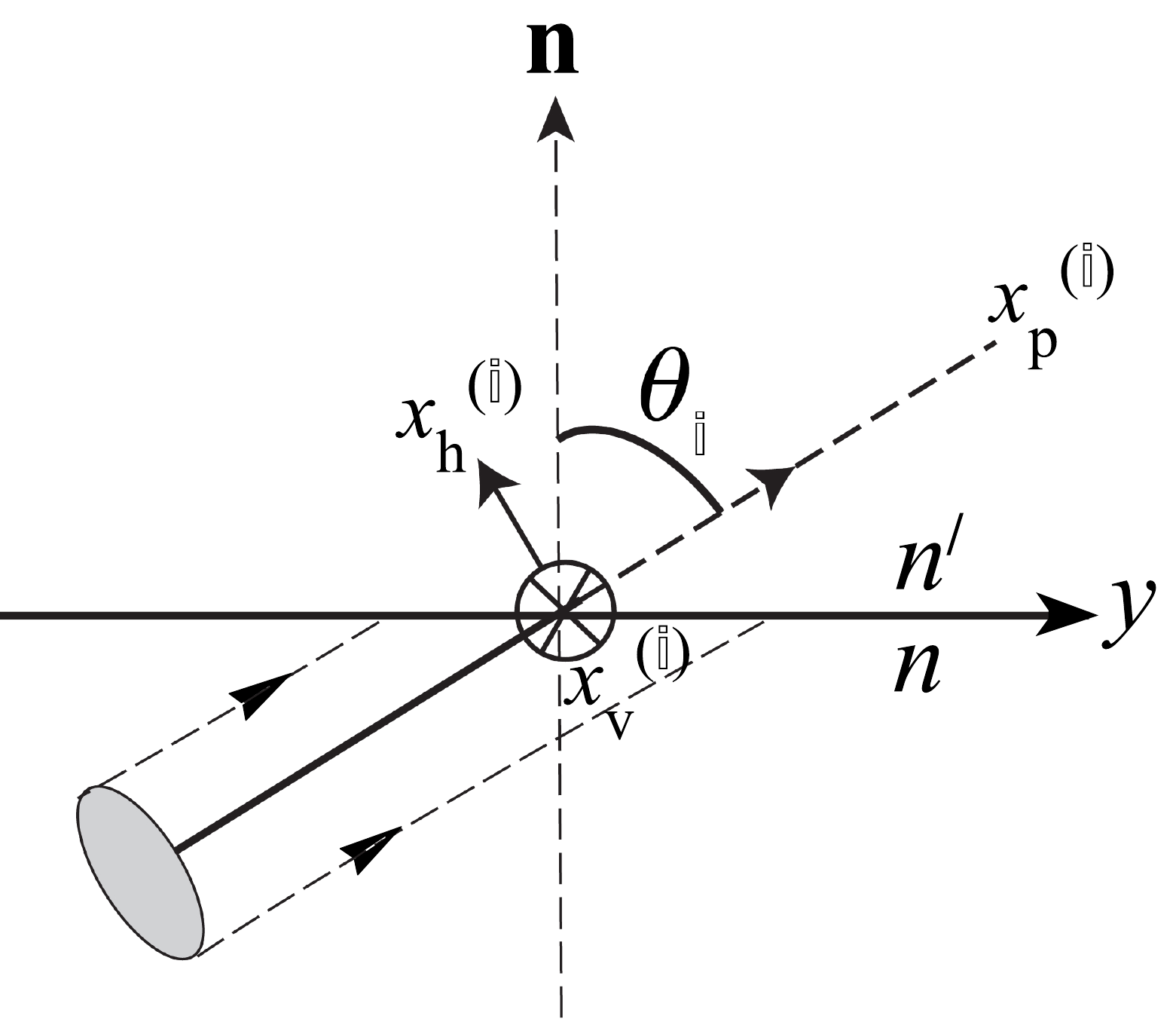}
  \qquad
  \caption{} \label{fig:in-beam-coord-sys}
    \end{figure}

\qquad Fig. \ref{fig:in-beam-coord-sys}. Illustrating the coordinate
system
$(x_{\text{v}}^{(\mathbb{i})},x_{\text{h}}^{(\mathbb{i})},x_{\text{p}}^{(\mathbb{i})})$
of the incident beam.

\begin{figure}[htbp]  \centering
  \includegraphics[scale=.8]{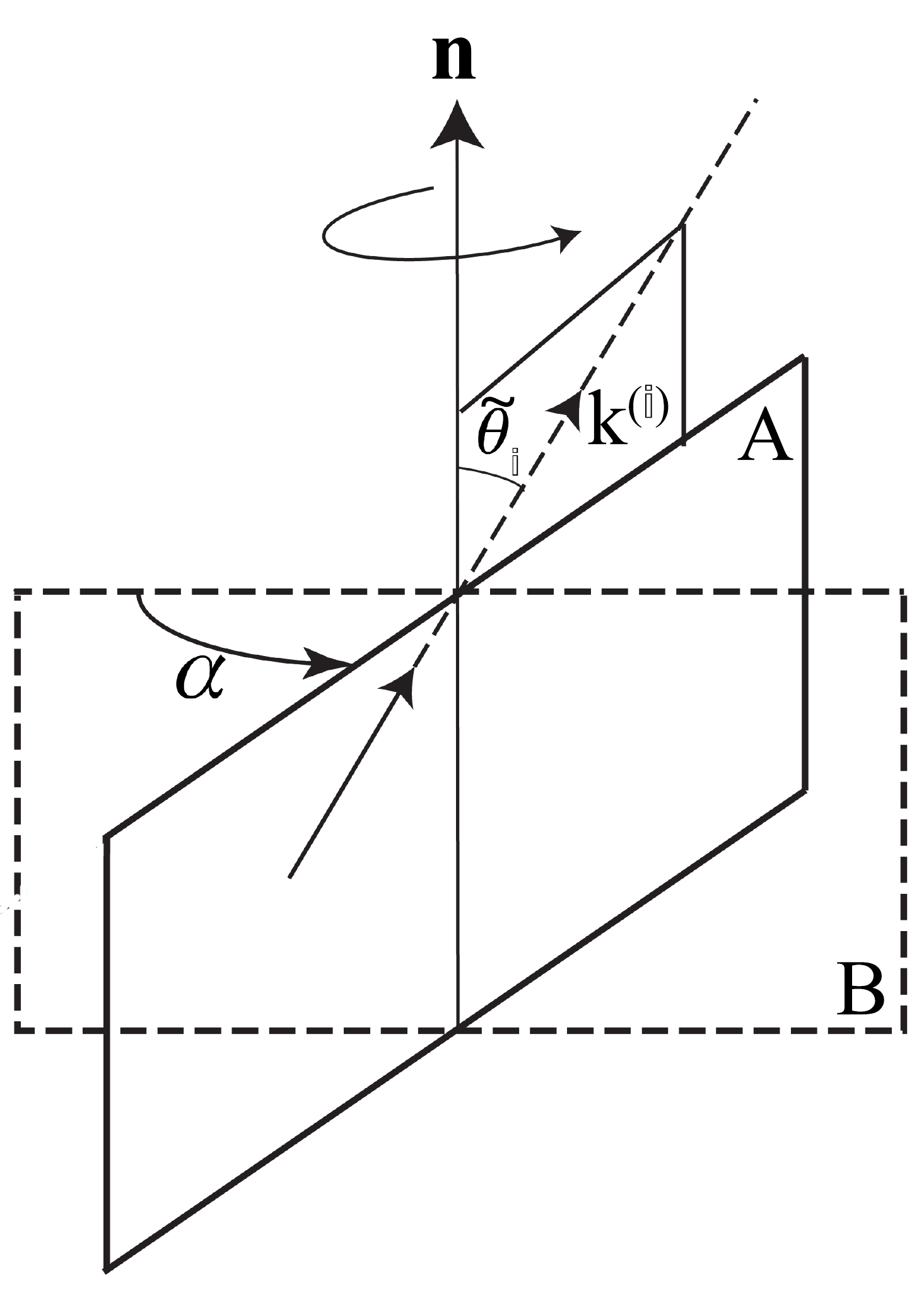}
  \qquad
  \caption{} \label{fig:xyz-z'y'z'-trans}
    \end{figure}

\qquad Fig. \ref{fig:xyz-z'y'z'-trans}. Illustrating the geometry
relating to the plane of incidence A of the plane wave component
with wave vector $\mathbf{k}^{(\mathbb{i})}$. The plane B indicated
in dotted line is the plane formed by the axis of the incident beam,
and the normal $\mathbf{n}$ to the interface.

\begin{figure}[htbp]  \centering
  \includegraphics[scale=.8]{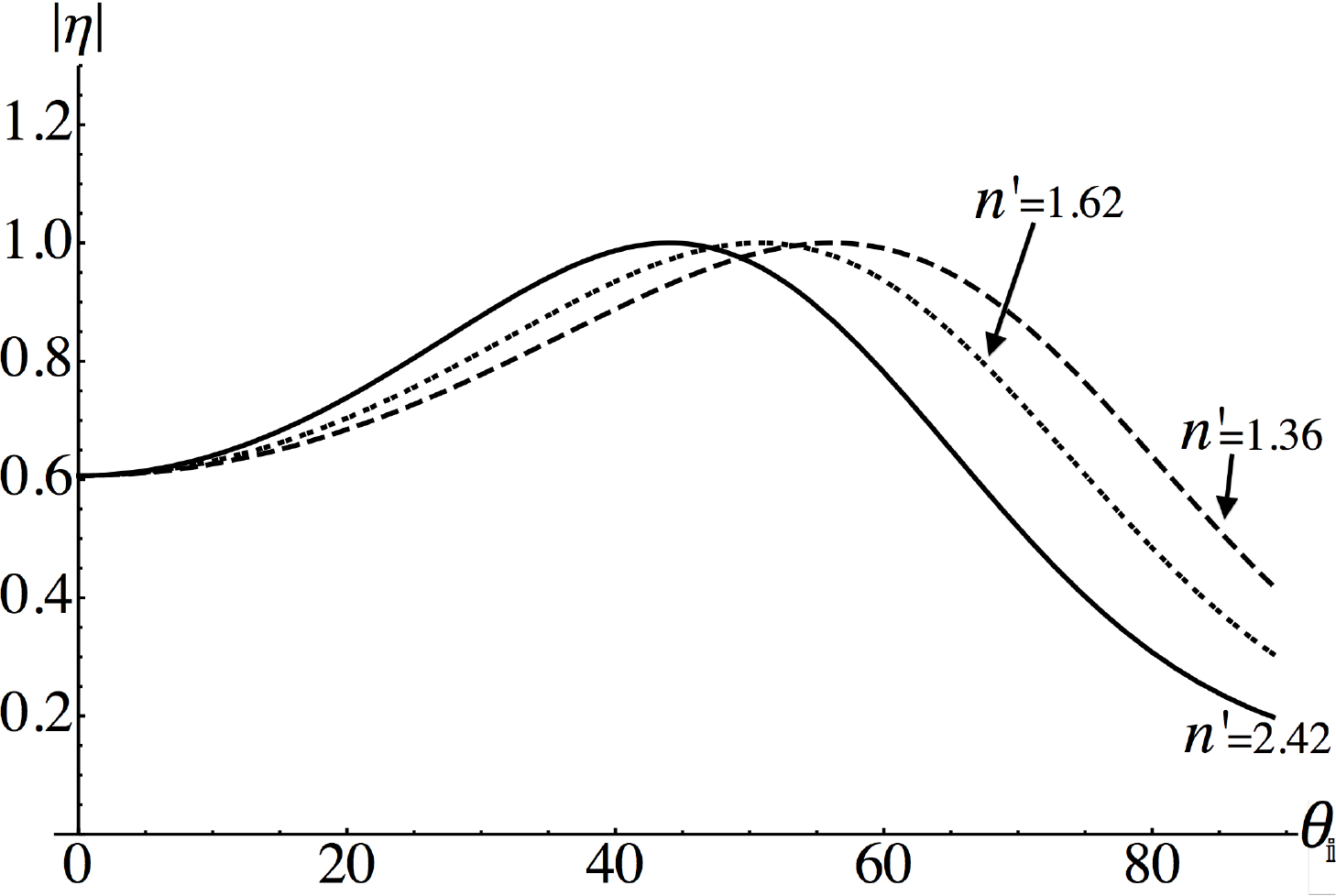}
  \qquad
  \caption{} \label{fig:trans-doc-th-dep}
    \end{figure}

\qquad Fig. \ref{fig:trans-doc-th-dep}. Modulus of the spectral
degree of coherence of the transmitted beam, at two fixed points on
the interface, plotted against the angle of incidence for ethanol
($n'=1.36$), for flint glass ($n'=1.62$) and for diamond
($n'=2.42$), at frequency $\omega \approx 3.2\times 10^{15}$
sec$^{-1}$, for the parameters $\delta=0.001$m, $\sigma=0.01$m,
$A_{\text{h}}/A_{\text{v}}=1$.

\begin{figure}[htbp]  \centering
  \includegraphics[scale=.8]{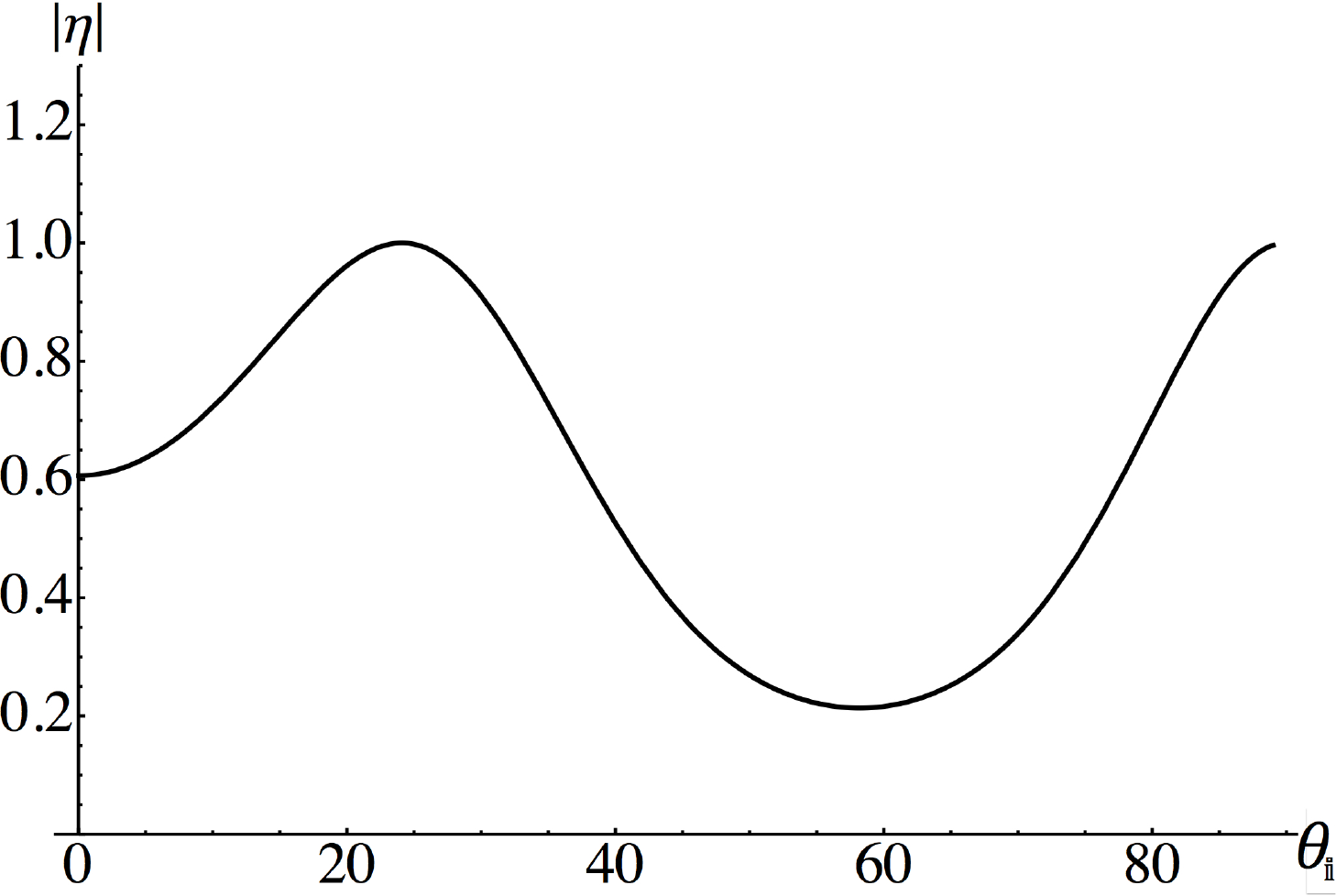}
  \qquad
  \caption{} \label{fig:refl-doc-th-dep}
    \end{figure}

\qquad Fig. \ref{fig:refl-doc-th-dep}. Modulus of the spectral
degree of coherence of the reflected beam, at two points located on
the interface, plotted against the angle of incidence, for ethanol
($n'=1.36$), for flint glass ($n'=1.62$), and for diamond
($n'=2.42$), for the same choice of parameters as used in Fig.
\ref{fig:trans-doc-th-dep}. All the three curves are identical.

\begin{figure}[htbp]
\centering
 \subfigure[~ethanol ($n'=1.36$)] {
    \label{subfig:doc-trans-th-56-89-ethanol}
     \includegraphics[scale=.65]{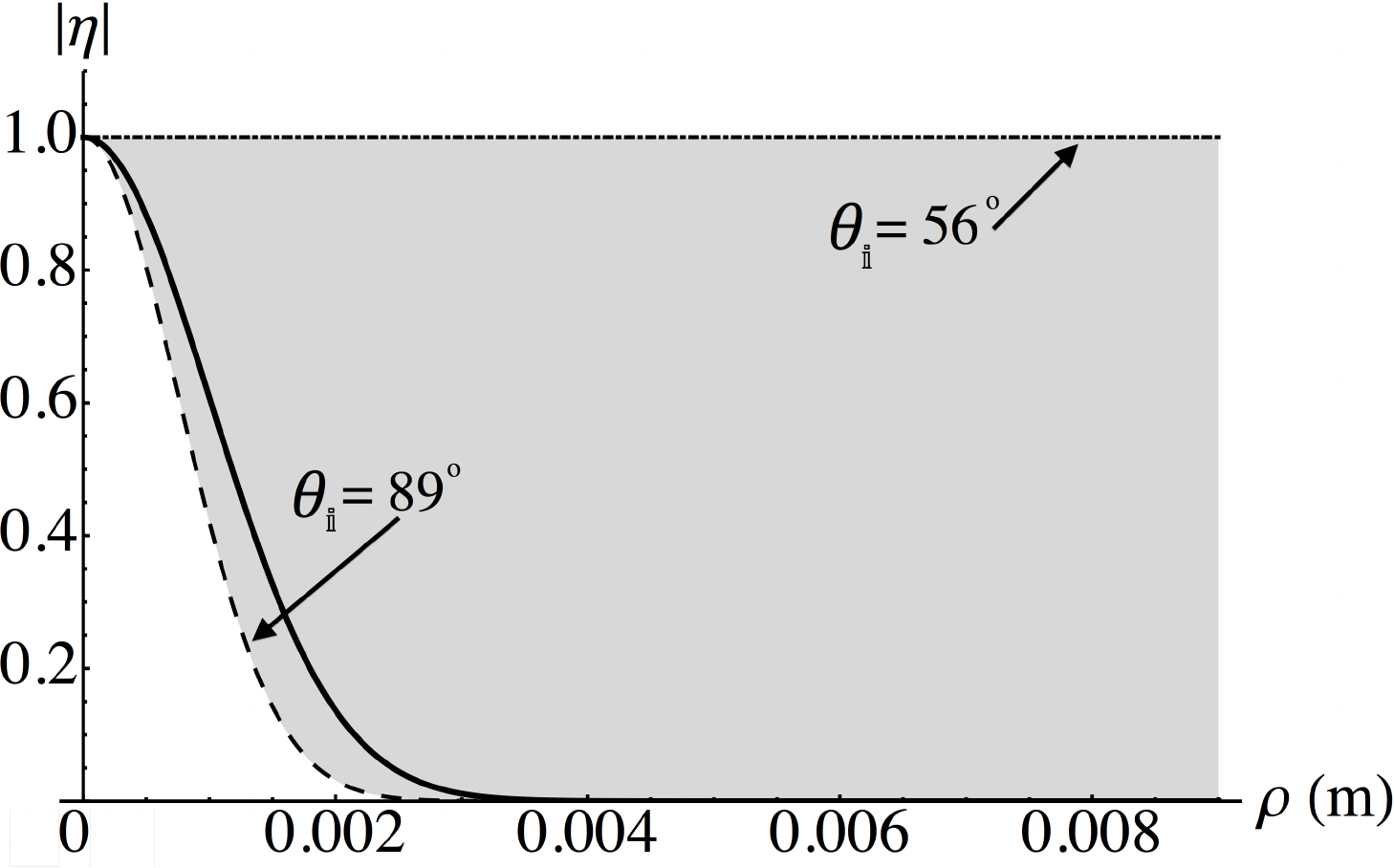}
} 
   \subfigure[~flint glass ($n'=1.62$)] {
    \label{subfig:doc-trans-th-50-89-flint-glass}
    \includegraphics[scale=.65]{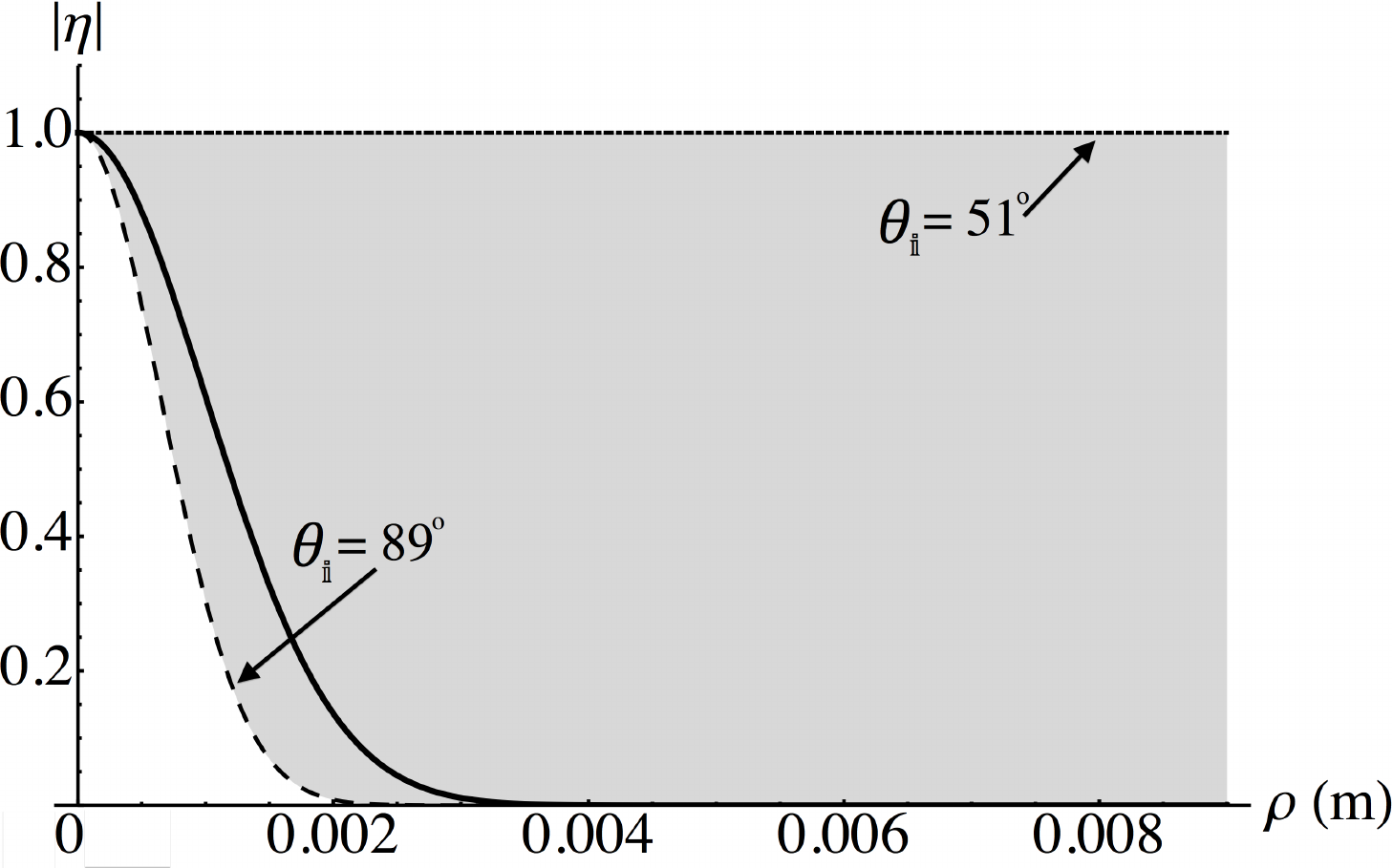}
} \subfigure[~diamond ($n'=2.42$)] {
    \label{subfig:doc-trans-th-44-89-diamond}
    \includegraphics[scale=.65]{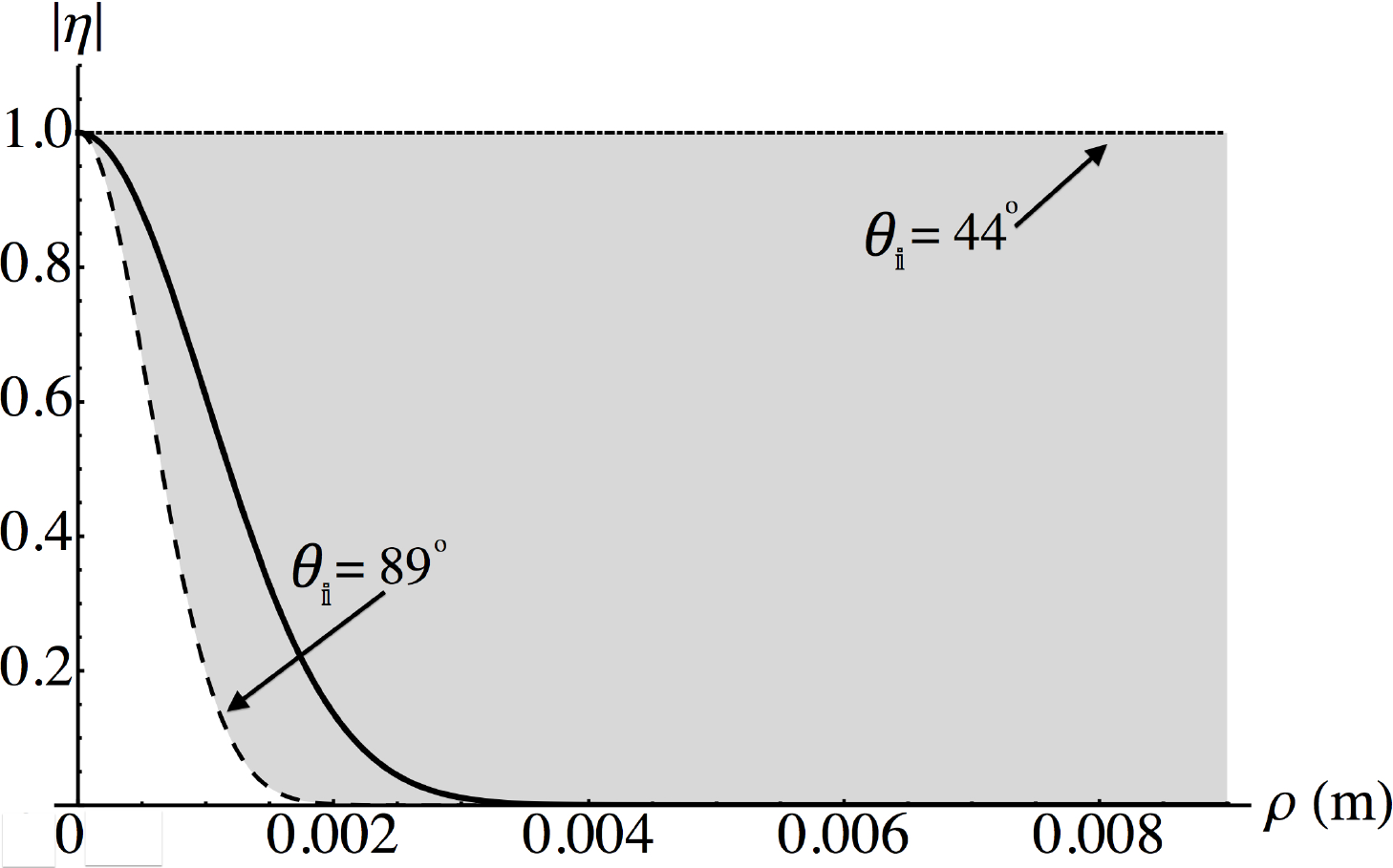}
}
\caption{} 
\label{fig:doc-trans-three-med}
\end{figure}

\qquad Fig. \ref{fig:doc-trans-three-med}. Modulus $|\eta|$ of the
spectral degree of coherence of the transmitted beam plotted as
function of $\rho=|\r'-\r|$, for two values of
$\theta_{\mathbb{i}}$, for ethanol (a), for flint glass (b) and for
diamond (c); the other parameters are same as before. The solid line
represent the modulus of the spectral degree of coherence at the
source plane. The shaded regions indicate the improvement and the
degradation of the degree of coherence that can be achieved on
transmission.

\begin{figure}[htbp]  \centering
  \includegraphics[scale=.95]{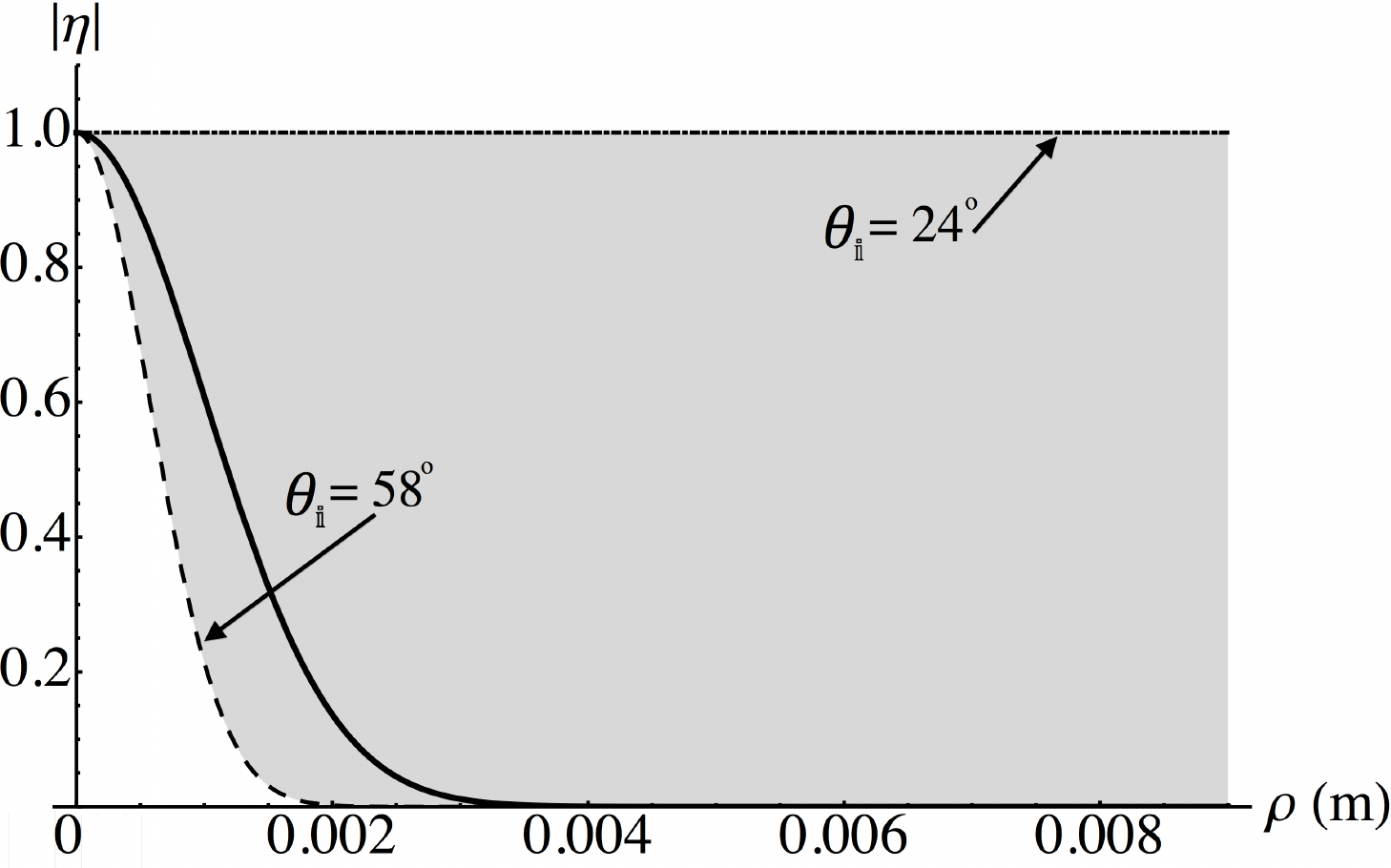}
  \qquad
  \caption{} \label{fig:doc-refl-th-25-50-flint-glass}
    \end{figure}

\qquad Fig. \ref{fig:doc-refl-th-25-50-flint-glass}. Modulus
$|\eta|$ of the spectral degree of coherence of the reflected beam
(same for all media) plotted as functions of $\rho=|\r'-\r|$ for
values of $\theta_{\mathbb{i}}=24^{\circ},58^{\circ}$, with the same
choice of the other parameters as used in Fig.
\ref{fig:trans-doc-th-dep}. The solid line represent the modulus of
the spectral degree of coherence at the source plane. The shaded
region indicates the improvement and the degradation of the degree
of coherence that can be obtained by reflection.

\newpage

\newpage

\newpage

\newpage

\newpage

\newpage

\newpage

\begin{thebibliography}{99}
\bibitem {Note-quant-coh-order} In the quantum theory of coherence
these correlations properties are referred to as first order ones.
\bibitem {Note-ang-spec} The formula is obtained from immediate
generalization of the scalar tratement presented in Ref. \cite{MW},
Sec. 5.6.
\bibitem {Note-tot-int-refl} In the case of the total internal
reflection, contributions from the plane wave components in the
transmitted field would be minimal, and therefore the effects due to
evanscent waves cannot be neglected.
\bibitem {Note-calc-illus} The treatment provided in
\cite{MW}, Sec. 5.6 is based on scalar theory, but it can be readily
generalized to vector theory for the case of optical beams, as we
did here for our purpose.
\end{thebibliography}
\end{document}